\def\nocolor{nocolor}
\ifx\scipostversion\undefined 
  \documentclass[aps,pra,twocolumn,superscriptaddress]{revtex4-1}
\else
  \documentclass[submission, Phys]{SciPost}
\fi

\ifx\scipostversion\undefined  
\fi   

\usepackage{graphicx}
\usepackage{amsmath}
\usepackage{bbold}
\usepackage[T1]{fontenc}
\usepackage{xcolor}
\usepackage{hyperref} 

\ifx\nocolor\undefined
   
   \newcommand{\tcolr}[1]{\textcolor{red}{#1}}
\else
   
   \newcommand{\tcolr}[1]{{#1}}
\fi

\newcommand{\beq}{\begin{equation}}
\newcommand{\eeq}{\end{equation}}
\newcommand{\beqs}{\begin{equation*}}
\newcommand{\eeqs}{\end{equation*}}
\newcommand{\beqa}{\begin{eqnarray}}
\newcommand{\eeqa}{\end{eqnarray}}
\newcommand{\beqas}{\begin{eqnarray*}}
\newcommand{\eeqas}{\end{eqnarray*}}
\def\bals#1\eals{\begin{align*}#1\end{align*}}
\def\bal#1\eal{\begin{align}#1\end{align}}

\newcommand{\bcent}{\begin{center}}
\newcommand{\ecent}{\end{center}}

\newcommand{\bitem}{\begin{itemize}}
\newcommand{\eitem}{\end{itemize}}

\newcommand{\phe}{\phantom{e}}
\newcommand{\phs}{\phantom{i}}

\newcommand{\ih}{i \hbar}

\makeatletter
\newcommand*\bt{\mathpalette\bt@{.7}}
\newcommand*\bt@[2]{\mathbin{\vcenter{\hbox{\scalebox{#2}{$\m@th#1\bullet$}}}}}
\makeatother

\makeatletter
\newcommand*\ct{\mathpalette\ct@{.7}}
\newcommand*\ct@[2]{\mathbin{\vcenter{\hbox{\scalebox{#2}{$\m@th#1\circ$}}}}}
\makeatother

\newcommand*{\covdev}{\text{\dh}}
\newcommand*{\covded}{\text{\it \dj}}

\begin{document}

\ifx\scipostversion\undefined 

\title{Manifold curvature and Ehrenfest forces with a moving basis}
             
\author{Jessica F. K. Halliday}
\affiliation{Theory of Condensed Matter,
             Cavendish Laboratory, University of Cambridge, 
             J. J. Thomson Ave, Cambridge CB3 0HE, United Kingdom}             

\author{Emilio Artacho}
\affiliation{Theory of Condensed Matter,
             Cavendish Laboratory, University of Cambridge, 
             J. J. Thomson Ave, Cambridge CB3 0HE, United Kingdom}
\affiliation{CIC Nanogune BRTA and DIPC, Tolosa Hiribidea 76, 
             20018 San Sebastian, Spain}
\affiliation{Ikerbasque, Basque Foundation for Science, 48011 Bilbao, Spain}

\begin{abstract}
  Known force terms arising in the Ehrenfest dynamics of quantum electrons 
and classical nuclei, due to a moving basis set for the former,
can be understood in terms of the curvature of the manifold hosting
the quantum states of the electronic subsystem. 
  Namely, the velocity-dependent terms appearing in the Ehrenfest forces 
on the nuclei acquire a geometrical meaning in terms of 
the intrinsic curvature of the manifold, 
while Pulay terms relate to its extrinsic curvature.
\end{abstract}

\date{\today}


\maketitle

\else   

\begin{center}{\Large \textbf{
Manifold curvature and Ehrenfest forces with a moving basis
}}\end{center}

\begin{center}
Jessica F. K. Halliday\textsuperscript{1},
Emilio Artacho\textsuperscript{1,2,3*}
\end{center}

\begin{center}
{\bf 1} Theory of Condensed Matter,
           Cavendish Laboratory, University of Cambridge, 
           J. J. Thomson Ave, Cambridge CB3 0HE, United Kingdom
\\
{\bf 2} CIC Nanogune BRTA and DIPC, Tolosa Hiribidea 76, 
           20018 San Sebastian, Spain
\\
{\bf 3} Ikerbasque, Basque Foundation for Science, 48011 Bilbao, Spain
\\
* ea245@cam.ac.uk
\end{center}

\begin{center}
\today
\end{center}


\section*{Abstract}
{\bf
  Known effects arising in the Ehrenfest dynamics of quantum electrons 
and classical nuclei with a moving basis set for the former,
can be understood in terms of the curvature of the manifold hosting
the quantum states of the electronic subsystem. 
  Namely, the velocity dependent terms appearing in the Ehrenfest forces 
on the nuclei acquire a geometrical meaning in terms of 
the intrinsic curvature of the manifold, 
while Pulay terms relate to its extrinsic curvature.
}

\fi  

\section{Introduction}

  In a recent paper \cite{Artacho2017} it was established how to reformulate 
quantum-mechanical equations for situations in which the basis set and the 
spanned Hilbert space vary with external parameters such as nuclear positions.
  This is routinely encountered in electronic structure calculations using
atom-centered basis functions, and where the nuclei move, that is,
any first-principles calculation method in quantum chemistry or 
condensed matter and materials physics using atomic orbitals as
basis sets. 
  There are many such methods and software packages that are widely used 
in either or both communities (for a brief review and links to codes used in 
quantum chemistry see, e.g.,  Ref.~\cite{Handbook2017}; for 
methods and programs using atomic orbitals in condensed matter 
see, e.g., Refs.~\cite{Garcia2020, Kuhne2020, Oliveira2020, 
Dovesi2020, QuantumATK2019, Blum2009, Delley2000}).
  We will restrict ourselves here to mean-field-like methods, such as 
Hartree-Fock or Kohn-Sham density-functional theory (DFT)
\cite{Kohn-Sham}, including their time-dependent versions 
\cite{Runge1984}, and we will therefore use single-particle 
language. 
  
  For adiabatic situations, it is an old and well-known problem, 
for which the relevant equations have long been established.
  The key consequence of the basis states moving with nuclei is 
the appearance of Pulay forces \cite{Pulay1969},
which are extra terms in the adiabatic forces acting on the nuclei due
to the moving basis.
  The generalisation to non-adiabatic problems was done two decades
ago \cite{Saalmann1996,Todorov2001,Kunert2003}.
  Differential geometry concepts were recently used to obtain a 
transparent formalism, which allowed better insights into the meaning of 
the extra terms appearing in the equations \cite{Artacho2017}.
  The use of the formalism was demonstrated in two examples,
namely, the time-dependent Schr\"odinger equation 
and the adiabatic forces on nuclei.

  In this paper we extend the formalism of Ref.~\cite{Artacho2017} 
to the Ehrenfest forces in mixed,
classical-nuclei / quantum-electrons dynamical calculations.
   We show that the extra force terms appearing beyond the Pulay 
forces consist of a term proportional to the Riemann curvature tensor 
of the fibre bundle and the nuclear velocities, and, therefore, explicitly 
non-adiabatic, plus a term depending on the connection, which is
implicitly non-adiabatic.
  The Pulay forces themselves are shown to appear only
for curved manifolds, including extrinsically curved.


\section{Forces} 

\subsection{General formalism}

For Ehrenfest dynamics, considering a system of quantum electrons 
and classical nuclei, and following Todorov~\cite{Todorov2001}, 
we start from the Lagrangian
\beq
\label{eq:Lagrangian}
L= \sum_n^{N_e} \langle \psi_n | i \hbar \, d_t - H | \psi_n \rangle 
+ \sum_j^{3N_n} \frac{1}{2} M_j \dot{R}_j^2 - V_{\mathrm{nuc}}(\{R_j\})
\eeq
defined for the $N_e$ wavefunctions $\psi_n$ of independent electrons
(we will disregard spin hereafter), 
\tcolr{and for the $3N_n$ position components of 
$N_n$ nuclei in three dimensions, $R_j$, as dynamical variables. 
  $M_j$ represent the nuclear masses; $j$ runs over all nuclear-position
vector components, and, therefore, the mass associated to the three 
components of a given nucleus is the same.} 
  $V_{\mathrm{nuc}}(\{R_j\})$ stands for  
the nucleus-nucleus repulsion.  
  $d_t$ represents the time derivative, $\frac{\mathrm{d}}{\mathrm{d}t}$,
indicating $|\dot{\psi}_n\rangle$ in the Lagrange sense as $d_t |\psi_n\rangle$
(to distinguish it from the Lagrange partial derivatives, 
$\partial_j = \frac{\partial}{\partial R_j}$ and $\frac{\partial}{\partial 
\langle \psi_n |}$),
although $d_t$ is still the partial time derivative in the Schr\"odinger
sense when referring to, for instance, $\psi_n(\mathbf{r},t)$.

  $H$ is the effective single-particle Hamiltonian for the electrons 
using a mean-field theory such as the Kohn-Sham version of
time-dependent density-functional theory \cite{Runge1984}.
  All the results of this work directly apply to that theory
\cite{Todorov2001, Kunert2003}.

The evolution of both $\psi_n$'s and $R_j$'s will be then defined 
by minimising the action $S=\int^t L \mathrm{d}t'$.
This evolution was shown to conserve total energy, total momentum 
and the orthonormality of the wavefunctions \cite{Todorov2001}.

Defining the first term as 
\begin{equation}
\label{eq:le}
L_e=\sum_n^{N_e} \langle \psi_n | i \hbar \, d_t - H | \psi_n \rangle \; ,
\end{equation}
we express now the electronic wavefunctions in a finite, non-orthogonal, 
and evolving basis set, $\{ |e_{\mu}, t\rangle, \, \mu = 1 \dots \mathcal{N} \} $, in an
evolving $\mathcal{N}$-dimensional Hilbert space $\Omega(t)$, 
\tcolr{always a subspace of the entire (ambient) Hilbert space $\mathcal{H}$.}
$\Omega(t)$ at all times defines a $(\mathcal{N}+1)$-dimensional 
fibre bundle $\Xi_t$.
In its natural representation \cite{Artacho2017}, and summing over 
repeated indices, Eq.~\eqref{eq:le} becomes
\begin{equation*}
L_e= i \hbar \, \psi_{n\mu} \covded_t  \psi^{\mu}_{\phs n} - 
\psi_{n\mu} H^{\mu}_{\phe\nu} \psi^{\nu}_{\phs n} \; ,
\end{equation*}
with 
\beqs
\psi^{\mu}_{\phs n} = \langle e^{\mu} | \psi_n\rangle \;  , \; \;
\psi_{n\mu} = \langle \psi_n | e_{\mu} \rangle \; , \; \;
H^{\mu}_{\phe\nu} = \langle e^{\mu} | H | e_{\nu} \rangle \, .
\eeqs
  The set $\{ |e^{\mu}, t\rangle, \, \mu = 1 \dots \mathcal{N} \}$ is the
dual basis of $\{|e_{\mu},t\rangle\}$, also a basis of $\Omega(t)$, 
satisfying 
$\langle e^{\mu},t |e_{\nu},t \rangle = \delta^{\mu}_{\phs\nu}, \, \forall \mu,\nu$
at any time $t$.
The symbol $\covded_t$ indicates the covariant time derivative 
in $\Xi_t$, defined as \cite{Artacho2017}
\beq
\label{eq:covariant}
\covded_t  \psi^{\mu}_{\phs n} = d_t \psi^{\mu}_{\phs n} 
+ D^{\mu}_{\phe\nu t} \psi^{\nu}_{\phs n} \; , 
\eeq
where $D^{\mu}_{\phe\nu t}  =  \langle e^{\mu} | d_t e_{\nu} \rangle$ 
gives the connection in the manifold (note the convention in the 
order of indices).

\tcolr{  There is also the possibility of orthonormalising the basis set
at each time by, for instance, a time-dependent L\"owdin transformation
from the original non-orthogonal basis. 
  In that case, the formalism remains, but it would simplify with the
vectors of the dual basis becoming identical with their direct-basis
 corresponding vectors, and the metric tensors becoming the 
identity matrix.
  The equations all stay as for the natural representation with no need
of distinguishing upper/lower (contravariant/covariant) indices.
  Numerically, however, it would be less efficient, so we keep the general
non-orthogonal formalism for generality.}

\subsection{Derivation of the forces}

  For the electrons, the Euler-Lagrange equations for the wavefunctions 
give \cite{Todorov2001} the time-dependent Schr\"odinger equation 
in the natural representation \cite{Artacho2017},
\beq
\label{eq:td-se}
H^{\mu}_{\phantom{e}\nu} \psi^{\nu}_{\phs n}=i\hbar \, \covded_t 
\psi^{\mu}_{\phs n} \; .
\eeq

For the evolution of the nuclear coordinates, the Euler-Lagrange equations 
on $R_j$
give
\begin{equation*}
M_j \ddot{R}_j = - \partial_j V_{\mathrm{nuc}}(\{R_l\}) + \partial_j L_e \; ,
\end{equation*}
the last term representing the Ehrenfest forces on the ions due to the electrons
(we will not include the nucleus-nucleus repulsion into the 
Ehrenfest forces as defined here).

  Let us assume henceforth that the time evolution of the basis is associated 
to the nuclear motion, such that $|e_{\mu},t\rangle = 
|e_{\mu},\{R_j(t)\}\rangle$.
  This assumption includes the most widely used moving bases, 
\tcolr{which are fixed-shape atomic orbitals $f(\mathbf{r})$ 
for one quantum particle in three-dimensional space ($\mathbf{r}$ is 
3D position) moving with the nuclei as $f(\mathbf{r}-\mathbf{v}t)$, 
being $\mathbf{v}$ the instantaneous velocity of the nucleus
a particular basis function moves with.}
  It is not limited to that case, however: the shape can vary
(and does not need to be atomic-like), as long as it depends
on atomic positions and not explicitly on time
\footnote{If the basis functions can vary in time for the same
atomic positions, the formalism would heve to be generalised,
with an extra time dimension in the bundle, and allowing for 
$\partial_t$ in addition to the $\partial_j$'s.}.
  We then define the $(\mathcal{N}+3N_n)$-dimensional 
fibre bundle $\Xi_R$ defined by $\Omega \left (\{R_j\}\right)$, as spanned 
by the basis $\{|e_{\mu},\{R_j\}\rangle\}$.
  The covariant derivative in this $\Xi_R$ manifold is now
\beq
\label{eq:covdev2}
\covdev_j  \psi^{\mu}_{\phs n} = \partial_j \psi^{\mu}_{\phs n} 
+ D^{\mu}_{\phe\nu j} \psi^{\nu}_{\phs n} \; , 
\eeq
with the corresponding connection,
$D^{\mu}_{\phe\nu j}  =  \langle e^{\mu} | \partial_j e_{\nu} \rangle$. 
  Both manifolds are related by any trajectory given by $\{R_j(t)\}$, which
implies $\covded_t = v_j \covdev_j$, being $v_j$ the nuclear velocities
\footnote{The basis vectors for the nuclear coordinates could also
be nonorthogonal, in which case the coordinates and velocities would go
as $R^j$ and $v^j$, respectively, but we will assume an orthonormal
basis here without loss of generality for the topic at hand}.

$L_e$ being a scalar, $\partial_j L_e=\covdev_j L_e$. 
We compute the Ehrenfest forces on the nuclei as 
\begin{equation*}
F_j = \covdev_j L_e = \covdev_j \left ( i \hbar \, \psi_{n\mu} \covded_t  
\psi^{\mu}_{\phs n} - \psi_{n\mu} H^{\mu}_{\phe\nu} \psi^{\nu}_{\phs n} 
\right ) \; .
\end{equation*}

We just need two other key facts to proceed. Firstly,
in the Euler -Lagrange equations, the $\psi$'s and the $R_j$'s are
treated as independent variables, which in the present formalism translates into 
$\covdev_j  \psi_{n\mu} =\covdev_j  \psi_{\phe n}^{\nu}=0$.

Secondly, using the Riemann curvature tensor of the bundle,
\beq
\label{eq:curv}
\Theta^{\mu}_{\phe j\nu k} \psi^{\nu}_{\phs n} \equiv \covdev_j \covdev_k 
\psi^{\mu}_{\phs n} - \covdev_k \covdev_j \psi^{\mu}_{\phs n} \; ,
\eeq
and the fact that $\covded_t = v_k \covdev_k$,
the double derivative in the first term of $F_j$ becomes
\beqs
\covdev_j \covded_t \psi^{\mu}_{\phs n} 
= v_k \, ( \covdev_k \covdev_j \psi^{\mu}_{\phs n}  + 
\Theta^{\mu}_{\phe j\nu k} \psi^{\nu}_{\phs n} ) 
=  v_k \, \Theta^{\mu}_{\phe j\nu k} \psi^{\nu}_{\phs n} \, ,
\eeqs
where we have used the fact that $\covdev_j \psi^{\mu}_{\phs n} =0$. 

These velocity-dependent terms in the forces
were amply discussed by Todorov \cite{Todorov2001}.
  To our knowledge, the fact that they are simply velocity times 
curvature was not known.
  The Ehrenfest forces can then be concisely written as
\beq
\label{eq:forces} \boxed{
F_j = - \, \psi_{n\mu} \left ( \covdev_j H^{\mu}_{\phe\nu} 
- i\hbar \, v_k \Theta^{\mu}_{\phe j\nu k}   \right )  \psi^{\nu}_{\phs n} .}
\eeq
By introducing the explicit expressions for the curvature 
$\Theta^{\mu}_{\phe j\nu k}$ and for the covariant derivative of the 
Hamiltonian $\covdev_j H^{\mu}_{\phe\nu}$ (both in 
Ref.~\cite{Artacho2017}), the resulting expression coincides
with what obtained in previous works by means of differential 
calculus \cite{Todorov2001,Kunert2003}.
  The meaning is now, however, much clearer, the expression more 
transparent, and the derivation much more direct.

\subsection{Non-adiabatic terms}

  In adiabatic (Born-Oppenheimer) evolution, the atomic forces 
can be expressed as \cite{Artacho2017}
\beq
\label{eq:bo-forces}
F_j^{BO} = - \, \psi_{n\mu} \, \partial_j H^{\mu}_{\phe\nu}  \, \psi^{\nu}_{\phs n} \; ,
\eeq
which is a direct result of the Hellmann-Feynman theorem.
  The difference between Eqs.~\eqref{eq:forces} and \eqref{eq:bo-forces}
gives the two non-adiabatic basis-related terms.

\subsubsection{Explicitly non-adiabatic}

  Firstly, the curvature term, 
$\psi_{n\mu} \ih \, v_k \Theta^{\mu}_{\phe j\nu k} \psi^{\nu}_{\phs n}$, 
is explicitly non-adiabatic, scaling linearly with the velocity of the displacing
basis functions, with the moving nuclei. 
\tcolr{
  In the adiabatic limit of atoms moving infinitely slowly, 
that force term vanishes with $v_j \rightarrow 0$.
  The curvature itself is still there in the adiabatic limit, and it
will give rise to effects analogous to the geometric phases
found in similar contexts~\cite{Bohm2003,Vanderbilt2018},
but the force itself is strictly non-adiabatic.} 

\tcolr{
  A possible visualisation of this force would be
that of a (generalised) centripetal force, which is suffered
by the nuclei due to the electrons being forced to evolve
within the curved manifold.
  A more canonical interpretation can be obtained from 
the close analogy of what is presented in this work and 
the theoretical framework of geometric phases in molecular 
and condensed matter physics (see~\cite{Bohm2003,Vanderbilt2018}
and references therein),
as was already noted in Ref.~\cite{Artacho2017}.
  In particular, the connection $D^{\mu}_{\phe\nu j}$
appearing in the covariant derivative of Eq.~\eqref{eq:covdev2} 
very closely relates to the Berry (or Mead-Berry) connection, or
gauge potential \cite{Bohm2003,Vanderbilt2018}, while the
curvature of Eq.~\eqref{eq:curv} relates to the corresponding
gauge field (gauge-covariant field strength).
  In this last sense, the velocity-dependent term in the
non-adiabatic force can be seen as a (generalised) 
Lorentz force for a charged particle in a magnetic field.
}

\tcolr{
  The fundamental difference should be kept in mind, however, 
between this paper's theory and that relating to the
mentioned geometric phases: the latter refers to the 
evolution of the problem's solutions, whereas the former
relates to the evolution of the basis set.
  As hinted in Ref.~\cite{Artacho2017}, the relation between
possible non-trivial behaviours in both manifolds could be 
interesting. 
  It is clear that a topologically non-trivial solution 
manifold can exist in a trivial basis manifold
(as in the limit of the latter tending to $\mathcal{H}$).
  The question is how non-trivial basis manifolds
affect the topology of the solutions manifold.
  To our knowledge it is still to be explored.
}

\subsubsection{Implicitly non-adiabatic}

  The second term is the difference between the
$\psi_{n\mu}  \, \covdev_j H^{\mu}_{\phe\nu} \,  \psi^{\nu}_{\phs n}$
term of Eq.~\eqref{eq:forces} and the 
$\psi_{n\mu}  \, \partial_j H^{\mu}_{\phe\nu} \,  \psi^{\nu}_{\phs n}$
term of Eq.~\eqref{eq:bo-forces}.
  Remembering \tcolr{the expression for the covariant derivative 
of the tensor associated to an operator,} $\covdev_j H^{\mu}_{\phe\nu} =
\partial_j H^{\mu}_{\phe\nu} + D^{\mu}_{\phe\sigma j} H^{\sigma}_{\phe\nu}
- H^{\mu}_{\phe\lambda} D^{\lambda}_{\phe\nu j}$ \tcolr{\cite{Artacho2017}}, 
that second non-adiabatic term can be expressed as 
\beq
\label{eq:diff}
\psi_{n\mu}  \, \left (
D^{\mu}_{\phe\sigma j} H^{\sigma}_{\phe\nu}
- H^{\mu}_{\phe\lambda} D^{\lambda}_{\phe\nu j}
 \right )  \psi^{\nu}_{\phs n} \; .
 \eeq
  This term is not explicitly vanishing with velocity, it is
rather an implicit non-adiabatic term. 
  This is seen from the Hellmann-Feynman theorem, 
whereby, in the adiabatic limit, $H^{\mu}_{\phe\nu} 
\psi^{\nu}_{\phs n} = E_n  \psi^{\mu}_{\phs n}$ and
$\psi_{n\mu} H^{\mu}_{\phe\nu} = E_n \psi_{n\mu}$.
Therefore
\bals
\psi_{n\mu}  \, & \left (
D^{\mu}_{\phe\sigma j} H^{\sigma}_{\phe\nu}
- H^{\mu}_{\phe\lambda} D^{\lambda}_{\phe\nu j}
 \right )  \psi^{\nu}_{\phs n} = \\
& =  \psi_{n\mu}  D^{\mu}_{\phe\sigma j} E_n \delta^{\sigma}_{\phe\nu}
\psi^{\nu}_{\phs n} 
- E_n \ \psi_{n\mu} \delta^{\mu}_{\phe\lambda} D^{\lambda}_{\phe\nu j}
 \psi^{\nu}_{\phs n}  \\
& = E_n  \, \psi_{n\mu}  \, \left (
D^{\mu}_{\phe\nu j}
- D^{\mu}_{\phe\nu j}
 \right )  \psi^{\nu}_{\phs n} = 0 \, .
 \eals

  However, if $\psi^{\mu}_{\phs n}$ is not an eigenstate of 
$H^{\mu}_{\phe \nu}$, i.e., it is evolving non-adiabatically, then
the term in  Eq.~\eqref{eq:diff} is not zero. 
\tcolr{
  Explicitly, given the basis of Hamiltonian eigenstates at 
any time $\xi^{\mu}_{\phs a}$ (that is,  
$H^{\mu}_{\phe\nu} \xi^{\nu}_{\phs a} = \varepsilon_a \xi^{\mu}_{\phs a}$),
if expanding the evolving $n$-th state as}
$\psi^{\mu}_{\phs n} = \sum_a C_{a n}  \xi^{\mu}_{\phs a}$,
Eq.~\eqref{eq:diff} becomes
\bals
\psi_{n\mu}  \, & \left (
D^{\mu}_{\phe\sigma j} H^{\sigma}_{\phe\nu}
- H^{\mu}_{\phe\lambda} D^{\lambda}_{\phe\nu j}
 \right )  \psi^{\nu}_{\phs n} = \\
& = \sum_{a,b} C^*_{an} C_{bn} (E_b -E_a) 
\xi_{a\mu}  D^{\mu}_{\phe\nu j}  \xi^{\mu}_{\phs b} \; ,
 \eals
  This expression becomes zero when at most only one of 
the $C_{an}$ coefficients is non-zero, which is precisely 
the adiabatic case.

\subsection{Pulay forces}

   There is one last point to make in the relation between forces
and the curvature of the manifold, which was already implicit in 
Ref.~\cite{Artacho2017}.
  Pulay forces appear in the calculation of the
matrix elements $\partial_i  H^{\mu}_{\phantom{e}\nu} 
= \partial_i \langle \psi^{\mu} | H | \psi_{\nu} \rangle$, as
\beq
\label{eq:pulay1}
\partial_i H^{\mu}_{\phantom{e}\nu} = 
\langle e^{\mu} | \partial_i H | e_{\nu} \rangle +
\langle \partial_i e^{\mu} | H | e_{\nu} \rangle + 
\langle e^{\mu} | H | \partial_i e_{\nu} \rangle \; , 
\eeq
the Pulay terms being the last two.
\tcolr{
  From the expression for the covariant derivative of
the Hamiltonian tensor, $\covdev_i H^{\mu}_{\phantom{e}\nu}$,
Ref.~\cite{Artacho2017} recast Eq.~\ref{eq:pulay1} in 
the more revealing form
}
\beqs
\covdev_i H^{\mu}_{\phantom{e}\nu} = 
\langle e^{\mu} |  \partial_i H | e_{\nu} \rangle +
\langle \partial_i e^{\mu} | Q_{\Omega} H | e_{\nu} \rangle +
\langle e^{\mu} | H Q_{\Omega} | \partial_i e_{\nu} \rangle ,
\eeqs 
where $Q_{\Omega}$ is  the complement of the projector
onto \tcolr{$\Omega\in\mathcal{H}$}, $P_{\Omega}$, i.e.,
$P_{\Omega} + Q_{\Omega} = \mathbb{1}$, the identity operator 
\tcolr{in the infinite-dimensional ambient Hilbert space $\cal{H}$.}
  The last two terms of the last expression make explicit
the direct relation between the Pulay correction and the 
curvature of the manifold: if the extrinsically defined basis
vectors stay within $\Omega$ when displacing coordinate $i$,
as would happen in the absence of curvature, 
\tcolr{
$Q_{\Omega} | \partial_i e_{\nu} \rangle = |0\rangle$ and
$\langle \partial_i e^{\mu} | Q_{\Omega}=\langle 0 |$, and,
therefore, the last two terms would be zero, giving 
$\covdev_i H^{\mu}_{\phantom{e}\nu} = 
\langle e^{\mu} |  \partial_i H | e_{\nu} \rangle$.} 
  The effect of basis change within $\Omega$ is taken care 
of by the connection in the covariant derivative in 
$\covdev_i H^{\mu}_{\phantom{e}\nu}$ [inside parenthesis in
Eq.~\eqref{eq:diff}],
which, as shown in the previous subsection, give a zero 
contribution to the forces in the adiabatic case. 
  In other words, when slightly displacing in nuclear configuration 
space, it is not the change in basis, but the turning of $\Omega$,
what matters for the (adiabatic) Pulay corrections to the forces, 
which happens when $\Xi_R$ is curved.

   Unlike the velocity-dependent terms discussed above, which
depend on the intrinsic (Riemann) curvature, 
the Pulay corrections appear for any curvature, including 
\tcolr{extrinsic curvature (in the sense of a cylinder having non-zero
extrinsic curvature but zero intrinsic one), since the Pulay corrections
stem from calculations in the ambient space $\mathcal{H}$, including
outside $\Omega$. The corrections will be there as long
as $Q_{\Omega} | \partial_i e_{\nu} \rangle \neq |0\rangle$.}


\section{Conclusion}

  For Ehrenfest dynamics of quantum electrons and classical 
nuclei, and for basis functions for the former that move with the latter,
it has been shown how the extra terms appearing in the
Ehrenfest forces acquire a natural geometric interpretation
in the curved manifold given by the set of electronic (tangent) 
Hilbert spaces defined at each set of nuclear positions. 
  The velocity-dependent term, explicitly non-adiabatic, depends on
the intrinsic curvature of the manifold (it could be considered
to be a centripetal force arising when constraining motion to
the curved manifold,
\tcolr{or the force arising due to the effective gauge field 
represented by that curvature}).
  It has the simple form of velocity times curvature.
  
  The two additional terms are implicitly non-adiabatic, disappearing
in the adiabatic limit, when following the Born-Oppenheimer surface.
  The well-known Pulay forces are also shown to be a consequence of
the manifold curvature, although in this case, an extrinsically
curved manifold is enough for these terms to appear.

  \tcolr{The paper allows a deeper understanding of the 
extra terms appearing in the Ehrenfest forces for moving
basis sets, in addition to connecting them to other contexts, 
albeit the forces themselves are unchanged.
  This is unlike what happens with the better understanding
of the electronic evolution equation, Eq.~\eqref{eq:td-se}, 
which enables the design of better numerical 
integrators~\cite{Halliday2021}.
   The curvature itself, however, Eq.~\eqref{eq:curv},
can also be exploited as a measure of basis incompleteness
along a nuclear trajectory.}

\ifx\scipostversion\undefined 
  \begin{acknowledgments}
\else
  \section*{Acknowledgments}
\fi
J. Halliday would like to acknowledge the EPSRC Centre for 
Doctoral Training in Computational Methods for Materials Science 
for funding under grant number EP/L015552/1.
  E. Artacho is grateful for discussions with Prof. Christos Tsagas, and 
acknowledges funding from the Leverhulme Trust, under Research 
Project Grant No. RPG-2018-254, from the EU through the 
ElectronStopping Grant Number 333813, within the Marie-Curie 
CIG program, and by  the Research Executive Agency under the 
European Union's Horizon 2020 Research and Innovation programme 
(project ESC2RAD, grant agreement no. 776410).
  Funding from Spanish MINECO is also acknowledged, through 
grant FIS2015-64886-C5-1-P, and from Spanish MICINN
through grant 
PID2019-107338RB-C61/ AEI /10.13039 / 501100011033.

\ifx\scipostversion\undefined 
  \end{acknowledgments}
\fi


\end{document}